\documentclass[aps,prl,twocolumn,amsmath,amssymb,floatfix,showkeys,showpacs]{revtex4-1}
\usepackage{bm} 
\usepackage{color}
\usepackage{graphics} 
\usepackage{graphicx} 
\usepackage{epsfig}
\usepackage{amssymb}
\usepackage{amsmath} 
\usepackage{amsthm}
\usepackage{amstext}
\usepackage{dcolumn} 
\usepackage{makeidx}
\usepackage{subfigure}

\begin{document}

\title{Kerr-like nonlinearities in an optomechanical system with an asymmetric anharmonic mechanical resonator}

\author{A.P. Saiko}
\email{saiko@physics.by}
\affiliation{Scientific-Practical Material Research Centre, Belarus National Academy of Sciences, 19 P. Brovka str., Minsk 220072 Belarus}

\author{R. Fedaruk}
\affiliation{Institute of Physics, University of Szczecin, 15 Wielkopolska str., 70-451, Szczecin, Poland}

\author{S.A. Markevich}
\affiliation{Scientific-Practical Material Research Centre, Belarus National Academy of Sciences, 19 P. Brovka str., Minsk 220072 Belarus}

\begin{abstract}
In the framework of the nonsecular perturbation theory based on the Bogoliubov averaging method, an optomechanical system with an asymmetric anharmonic mechanical resonator is studied. The cross-Kerr interaction and the Kerr-like self-interaction of photons and vibration quanta arise in the Hamiltonian. These interactions are induced by both cubic and quartic nonlinearities of oscillations of the mechanical resonator and the cavity-resonator interaction that is linear in mechanical displacements. We demonstrate a bistable behavior of the number of vibration quanta and find that this behavior is controlled by the cross-Kerr interaction. It is shown that, without driving and dissipation, the constructed superposition Yurke-Stoler-like states of the cavity (or the mechanical resonator) disentangle at certain times the entangled modes of the system. The obtained results offer new possibilities for control of optomechanical systems with asymmetric mechanical oscillations.
\end{abstract}

\maketitle
Optomechanical systems offer the possibility for control of light by mechanical motion and vice versa. The coupling between light and mechanical resonator vibrations is usually achieved via light pressure. The coupling can change the resonant frequency of the mechanical resonator and its damping. The latter can be used for cooling \cite{p1, p2, p3} or amplification \cite{p4}. The nonlinearity of the optomechanical interaction enables the realization of quantum squeezed states \cite{p5}. These states may be created only in the strong-coupling regime \cite{p6}, where a coupling constant is larger than the decay rates of the cavity and the mechanical resonator. The multi-photon strong coupling regime is accessible under a strong driving field on the cavity mode and the majority of the observed physical phenomena can be understood using a linear description \cite{p6}. In this case, nonclassical states \cite{p7, p8}, quantum entanglement \cite{p9, p10, p11}, quantum state transfer \cite{p12, p13, p14}, optomechanically induced transparency \cite{p15, p16, p17}, and normal-mode splitting \cite{p18, p19} have been investigated. The optomechanical interaction is intrinsically nonlinear \cite{p6}. In experiments this nonlinearity so far has played only a role in the classical regime of large amplitude light and mechanical oscillations.

The cavity-resonator interaction can be enhanced, for example, via the non-linearity of the Josephson effect \cite{p20}. This non-linearity leads to an additional nonlinear interaction, namely, a cross-Kerr coupling between the cavity and the resonator. The higher-order interactions in the displacement have also been investigated in devices with special geometry of the resonator \cite{p21, p22, p23}. The cross-Kerr coupling induces a change to the refractive index of the cavity depending on the number of vibration quanta in the resonator, whereas the radiation pressure coupling gives rise to the Kerr effect depending on the displacement of the resonator. In nonlinear optics \cite{p24}, the Kerr effect usually appears in nonlinear dispersive media due to third-order matter-light interactions. It was shown \cite{p25} that the nonlinear Kerr Hamiltonian can be formulated explicitly using the Born-Oppenheimer approximation for a standard Hamiltonian of optomechanical system (in the frame rotating with the driving field frequency). The same result was obtained using polaron-like transformation \cite{p26, p27}.

Mechanical resonators of optomechanical systems are usually modeled by harmonic oscillators or rarely by nonlinear Duffing oscillators with fourth-order nonlinearity, which generally is very weak. However, nonlinearities of mechanical resonators can significantly be increased by technological improvements of used materials, geometry of the system and additional adaptations \cite{p28, p29, p30}. For mechanical resonators, the role of asymmetric potentials, which are non-invariant under reversal of displacement sign, has been less studied. We mean oscillators with potentials including linear and cubic terms in displacement. Optomechanical systems, in which the mechanical resonator is modeled by the anharmonic oscillator with the Coulomb-interaction-dependent forcing terms, have been considered in \cite{p31, p32, p33}. In the frame of this model in optomechanics, the possibility of precision measuring electrical charge with optomechanically induced transparency \cite{p31}, Coulomb-interaction-dependent effect of high-order sideband generation \cite{p32} as well as force-induced transparency and conversion between slow and fast lights \cite{p33} have been studied. Recently, the electromagnetically induced transparency with a cubic nonlinear movable mirror has been considered \cite{p34}. Steady-state mechanical squeezing via Duffing and cubic nonlinearities was analyzed \cite{p35}. Probing Duffing and cubic anharmonicities of quantum oscillators in an optomechanical cavity was also studied in \cite{p36}.

In the present paper, in the framework of the nonsecular perturbation theory based on the Bogoliubov averaging method, we study an optomechanical system with an asymmetric anharmonic mechanical resonator. Due to the averaging, anharmonic mechanical oscillations, cubic and quartic in displacements, together with the cavity-resonator interaction, linear in displacements, result in the Kerr and cross-Kerr nonlinearities which are determined features of the optomechanical system.

The basic physics of the optomechanical system can be captured in the following Hamiltonian \cite{p6}:
\begin{equation}\label{eq1}
H=H_{0} +V,
\end{equation}
\[H_{0} =\omega _{c} a^{\dag } a+\Omega b^{\dag } b, \, \, \, \, V=-ga^{\dag } a(b^{\dag } +b), \]
where $\omega _{c} $ is the cavity frequency, $\Omega $ is the mechanical frequency, $g$ is the optomechanical coupling, and $a$ ($b$) represents the cavity's photon (vibration quantum) annihilation operator (we take the Planck constant $\hbar =1$). Oscillations of the mechanical resonator in a potential with weak cubic and quartic terms in displacements are taken into account by
\begin{equation}\label{eq2}
V_{anh} =\frac{1}{6} v(b+b^{\dag } )^{3} +\frac{1}{6} w(b+b^{\dag } )^{4} ,
\end{equation}
where $v=v^{0} (1/2m\Omega )^{3/2} $, $w=w^{0} (1/2m\Omega )^{2} $, $m$ is the mass of mechanical resonator, $v^{0} $ and $w^{0} $ are parameters describing the values of cubic and quartic anharmonicities, respectively. These anharmonicities can be realized by specific construction of the mechanical resonator. For example, the mechanical resonator in \cite{p31, p32, p33} is modeled by the anharmonic oscillator with the Coulomb-interaction-dependent forcing terms. In the interaction representation, nonlinear terms are given by
\[e^{iH_{0} t} (V+V_{anh} )e^{-iH_{0} t} =-ga^{\dag } a(b^{\dag } e^{i\Omega t} +H.c.)+\]
\begin{equation}\label{eq3}
+\frac{1}{6} v(b^{\dag } e^{i\Omega t} +H.c.)^{3} +\frac{1}{6} w(b^{\dag } e^{i\Omega t} +H.c.)^{4} \equiv H_{int} (t).
\end{equation}
Since for real optomechanical systems inequalities $\Omega {\rm \gg }g, v, w$ are well fulfilled, we can use the non-secular perturbation theory for averaging fast oscillations $e^{\pm in\Omega t} $ (where $n=1, 2, 3, 4$) in the time-dependent interaction Hamiltonian (3) and to obtain an approximately diagonal or diagonal time-independent effective Hamiltonian. In the canonical form it can be realized using the Bogoliubov averaging method \cite{p37, p38, p39}. Averaging up to the second order in small parameters $g/\Omega $, $v/\Omega $, $w/\Omega $ we obtain
\[H_{int} \to H_{int}^{_{eff} } =H_{int, 1}^{eff} +H_{int, 2}^{eff} , \]
where
\[H_{int, 1}^{eff} =<H_{int} (t)>, \]
\begin{equation}\label{eq4}
H_{int, 2}^{eff} =\frac{i}{2} \langle [\int^t d\tau (H_{int} (\tau )-<H_{int} (\tau )>), H_{int} (t)]\rangle .
\end{equation}

Here the symbol $\left\langle ...\right\rangle $ denotes time averaging over rapid oscillations of the type $e^{\pm in\Omega t} $ given by $\langle O(t)\rangle =\frac{\Omega }{2\pi } \int _{0}^{2\pi /\Omega } O(t)dt$ and the upper limit \textit{t} of the indefinite integral indicates the variable on which the result of the integration depends, and square brackets denote the commutation operation.

Calculations based on Eq. ~\ref{eq4} give
\[H_{int, 1}^{eff} =w(b^{\dag } b+b^{\dag } bb^{\dag } b), \]
\[H_{int, 2}^{eff} =\frac{gv}{\Omega } a^{\dag } a-\frac{5v^{2} }{6\Omega }b^{\dag } b-\]
\begin{equation}\label{eq5}
-\frac{g^{2} }{\Omega } a^{\dag } aa^{\dag } a+\frac{2gv}{\Omega } a^{\dag } ab^{\dag } b-\frac{5v^{2} }{6\Omega } b^{\dag } bb^{\dag } b.
\end{equation}
Contributions proportional to $(w/\Omega )^{2} $, $vw/\Omega ^{2}$, and $gw/\Omega ^{2}$ are neglected.

In the laboratory frame we add the driving term for the cavity $H_{d, a} =i\varepsilon (a^{\dag } e^{-i\omega _{d} t} -H.c.)$ and the driving term for the mechanical resonator $H_{d, b} =i\eta (b^{\dag } e^{-i\Omega _{d} t} -H.c.)$, where $\varepsilon $ and $\eta $ are the amplitudes of photonic and vibration driving fields. Using evolution operators
${U_a} = {e^{i{\omega _d}{a^\dag }at}}$ and ${U_b} = {e^{i{\Omega _d}{b^\dag }bt}}$, the effective Hamiltonian in the rotating frame can be written as
\begin{equation}\label{eq6}
H^{eff} =H_{0}^{eff} +H_{1}^{eff} +H_{d} ,
\end{equation}
\[H_{0}^{eff} =\tilde{\omega }_{c} a^{\dag } a+\tilde{\Omega }b^{\dag } b, \]
\[H_{1}^{eff} =-\chi _{a} a^{\dag } aa^{\dag } a+\chi _{ab} a^{\dag } ab^{\dag } b-\chi _{b} b^{\dag } bb^{\dag } b, \]
\[H_{d} =i\varepsilon (a^{\dag } -a)+i\eta (b^{\dag } -b), \]
 where $\Delta =\omega _{c} -\omega _{d} $ , $\delta =\Omega -\Omega _{d} $, $\tilde{\omega }_{c} =\Delta +gv/\Omega $, $\tilde{\Omega }=\delta -5v^{2} /6\Omega +w$, $\chi _{a} =g^{2} /\Omega $, $\chi _{b} =5v^{2} /6\Omega -w$, and $\chi _{ab} =2gv/\Omega $.

 The Hamiltonian $H_{0}^{eff} $ represents the cavity in the rotating frame and the mechanical resonator with renormalized frequencies due to cubic and quartic anharmonicities. The first term in $H_{1}^{eff}$ describes the Kerr interaction of photons in the cavity; the second term represents the cross-Kerr interaction of photons and vibration quanta induced by interference contribution of both the cubic nonlinearity of oscillations of the mechanical resonator and the cavity-resonator interaction linear in mechanical displacements; the third term describes the Kerr-like mechanical self-interaction of the resonator. Hence, the back-action of linear oscillations of the mechanical resonator in the cavity results in the Kerr effect for the cavity field $\chi _{a} a^{\dag } aa^{\dag } a$, with $\chi _{a} $ the Kerr frequency shift per photon. At the same time, even weak asymmetric anharmonicities induce the cross-Kerr effect of the cavity and the resonator $\chi _{ab} a^{\dag } ab^{\dag } b$, with $\chi _{ab} $ the cross-Kerr frequency shift per photon or per vibration quantum. The frequency shifts due to the induced cross-Kerr effect depend on the number of photons in the cavity and of vibration quanta in the resonator.

Taking into account driving and dissipation, the quantum master equation for the density matrix $\rho $ of our system is
\begin{equation}\label{eq7}
 \frac{d\rho }{dt} =-i[H^{eff} , \rho ]+L_{c} \rho +L_{m} \rho .
\end{equation}

The incoherent coupling of the system with its environment is modeled \cite{p40} by the Lindblad dissipators
\[{L_c}\rho  = \frac{\kappa }{2}(2a\rho {a^\dag } - {a^\dag }a\rho  - \rho {a^\dag }a),\]
\[L_{m} \rho =\frac{\gamma }{2} (\bar{N}+1)(2b\rho b^{\dag } -b^{\dag } b\rho -\rho b^{\dag } b)+\]
\begin{equation}\label{eq8}
+\frac{\gamma }{2} \bar{N}(2b^{\dag } \rho b-bb^{\dag } \rho -\rho bb^{\dag } ).
\end{equation}
A photon that has entered the cavity decays at a rate $\kappa $ either by transmission or due to absorptive losses inside the cavity. The decay of mechanical energy is characterized by a damping rate $\gamma $. We assume here a zero-temperature bath for the optical cavity and a mean thermal occupation $\overline N  = {\left[ {\exp (\Omega /{k_B}T) - 1} \right]^{ - 1}}$ of the mechanical resonator bath at the frequency $\Omega $. Using Eqs. ~\ref{eq6} \t ~\ref{eq8}, the equations of motion for mean values of dynamical variables in the frame rotating with the driving field frequency can be written as
$$\frac{d\langle a\rangle }{dt} =-i(\tilde{\omega }_{c} -\chi _{a} )\langle a\rangle -\frac{\kappa }{2} \langle a\rangle +i2\chi _{a} \langle a^{\dag } aa\rangle -i\chi _{ab} \langle ab^{\dag } b\rangle +\varepsilon, $$

\begin{equation}\label{eq9}
\frac{d\langle b\rangle }{dt} =-i(\tilde{\Omega }-\chi _{b} )\langle b\rangle -\frac{\gamma }{2} \langle b\rangle +i2\chi _{b} \langle b^{\dag } bb\rangle -i\chi _{ab} \langle a^{\dag } ab\rangle +\eta .
\end{equation}

As an example, consider the semiclassical approximation $\langle a\rangle =\alpha $, $ \langle a^{\dag } aa\rangle =\left|\alpha \right|^{2} \alpha $, $ \langle b\rangle =\beta $, $ \langle b^{\dag } bb\rangle =\beta \left|\beta \right|^{2} $, $ \langle ab^{\dag } b\rangle =\alpha \left|\beta \right|^{2} $. In the steady state we have the following self-consistent system of equations for determining the mean number of photons $ \bar{n}_{a} =\left|\alpha \right|^{2} $ in the cavity (the mean number of vibration quanta $ \bar{n}_{b} =\left|\beta \right|^{2} $ in the mechanical resonator) as a function of the power of the photonic and vibration driving fields, detunings $ \delta $ and $ \Delta $ as well as the Kerr and cross-Kerr couplings:
\begin{equation}\label{eq10}
\left[\frac{\gamma ^{2} }{4} +(\tilde{\Omega }-\chi _{b} -2\chi _{b} \bar{n}_{b} +\chi _{ab} \bar{n}_{a} )^{2} \right]\bar{n}_{b} =\eta ^{2} ,
\end{equation}
\begin{equation}\label{eq11}
\left[\frac{\kappa ^{2} }{4} +(\tilde{\omega }_{c} -\chi _{a} -2\chi _{a} \bar{n}_{a} +\chi _{ab} \bar{n}_{b} ){}^{2} \right]\bar{n}_{a} =\varepsilon ^{2} .
\end{equation}

Now we analyze solutions of Eqs. (10,11) for parameters wich can be realized, for example, in the device with optomechanical coupling between a multilayer graphene mechanical resonator and a superconducting microwave cavity \cite{p41}.
The mean number $\bar{n}_{b} $ of vibration quanta in the mechanical resonator versus the amplitude of the vibration driving field and the cross-Kerr parameter is presented in a counter plot (Fig. 1 (b)). The cross-Kerr parameter $ \chi _{ab} $ can be positive or negative depending on the sign of the parameter of cubic anharmonicity. The vertical lines \textit{1} - \textit{4} in Fig. 1 (b) show cuts at the values of $ \chi _{ab} $ for which the dependences of $ \bar{n}_{b} $ on the normalized amplitude $ \eta /\gamma $ of the vibration driving field (Fig. 1 (a)) were obtained. To illustrate hysteresis, arrows depict jumps in $ \bar{n}_{b} $ when the driving field increases or decreases. One can see that the range of the amplitudes of the vibration driving field, for which the bistable behavior of $ \bar{n}_{b} $ can exist, is larger at the negative cross-Kerr parameter  than at the positive one (Fig. 1 (a)). Moreover, increasing the positive cross-Kerr parameter $ \chi _{ab} $ results in crossover from the bistable to monotonic (the line \emph{1} in Fig. 1 (a)) behavior. The dependence of $ \bar{n}_{b} $ on the mechanical driving detuning $ \delta $ from the resonator frequency also demonstrates the bistability (Fig. 1 (c)). An increase of the positive cross-Kerr parameter decreases the bistability area. The obtained hysteresis dependences of ${\bar n_b}(\eta )$   are similar to those for the Duffing oscillator. However, there is an important distinguishing feature such as the cross-Kerr contribution, ${\chi _{ab}}{\bar n_a}(\varepsilon )$, which is bistable due to the bistable behavior of  ${\bar n_a}(\varepsilon )$. Therefore, the ${\bar n_b}(\eta )$   dependence is determined not only by  the strength of the photon driving field  $\varepsilon$. It also strongly depends on which the stable or metastable branch in the bistable behavior of  ${\bar n_a}(\varepsilon )$ is selected. The stable branch of ${\bar n_a}(\varepsilon )$ was chosed in the culculations presented in Fig. 1.

\begin{figure}[ht]
\center{\includegraphics[width=1\linewidth]{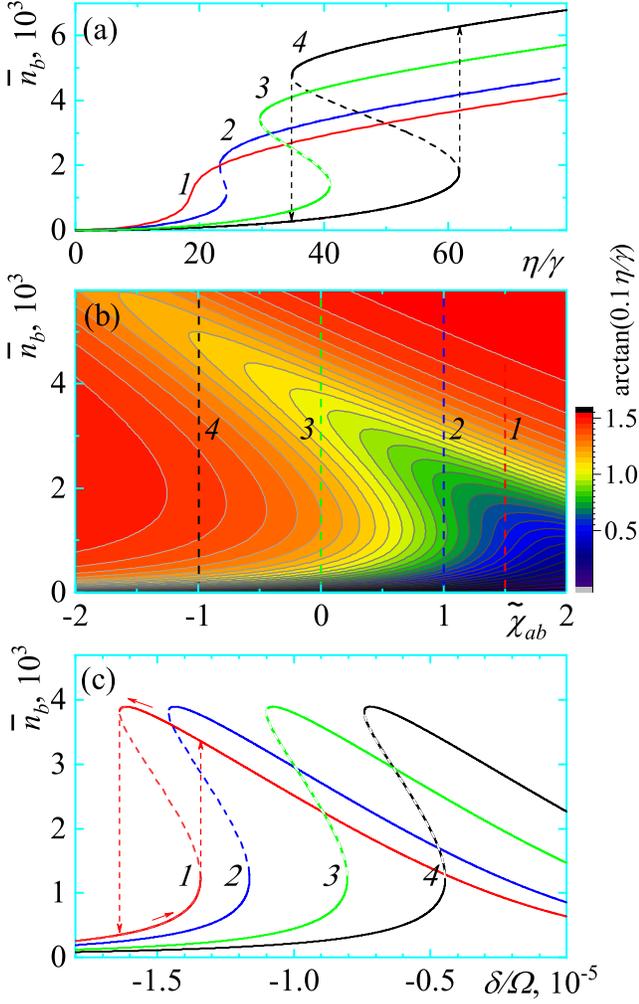}}
\caption{Fig. 1. The mean number $ \bar{n}_{b} $ of vibration quanta in the mechanical resonator. (a) $ \bar{n}_{b} $ versus the normalized amplitude $ \eta /\gamma $ of the vibration driving field for four values of the normalized cross-Kerr parameter $ \tilde{\chi }_{ab} =\chi _{ab} /4.59\times 10^{-8} Hz $ : \textit{1}\t 1.5, \textit{2}\t 1, \textit{3}\t 0, and \textit{4}\t -1. The parameters are $ \Omega /2\pi =36.2 $ \, MHz, $ g/2\pi =0.83 $ \, Hz, $ \kappa /2\pi =242 $ \, kHz, $ \gamma /2\pi =228 $ \, Hz, $ v/2\pi =1.0 $ \, Hz, $ w/2\pi =0.05 $ \, Hz, $ \Delta =1\times 10^{-2} \Omega $, $ \delta = -1\times10^{-5} \Omega $, $ \varepsilon /\kappa =8.39\times 10^{4} $, and $ \bar{n}_{a} =2.83\times 10^{9} $. (b) $ \bar{n}_{b} $ versus $ \eta /\gamma $ and $ \tilde{\chi }_{ab} $. The vertical lines show the values of $ \tilde{\chi}_{ab} $ presented in (a) and (c). (c) $ \bar{n}_{b} $ as a function of the mechanical driving detuning $ \delta $ from the resonator frequency at $ \eta /\gamma =31.22 $. The other parameters in (b) and (c) are the same as in (a).}
\end{figure}

Let us consider the evolution of the system when there are no driving terms in Hamiltonian ~\ref{eq6} ($\varepsilon , \eta =0$) and no dissipation in Eq. ~\ref{eq8} ($ \gamma , \kappa =0$). In this case the effective Hamiltonian is diagonal. It means that there is no exchange by quanta between the cavity and the mechanical resonator. From Eq. ~\ref{eq6}, one can see that $[H_{0}^{eff} , H_{1}^{eff} ]=0, $ thus, we obtain in the interaction representation: $ H^{eff} \to e^{iH_{0}^{eff} t} H_{1}^{eff} e^{-iH_{0}^{eff} t} =H_{1}^{eff} .$ When the cavity mode is initially in the vacuum state and the vibration mode is in the coherent state $ {\left| \beta \right\rangle}$, two terms in $ H_{1}^{eff}$ containing $ a^{\dag } a $ and $ (a^{\dag } a)^{2} $ do not contribute, and the evolution of the system is determined by the Kerr term $ -\chi _{b} (b^{\dag } b)^{2} . $ Therefore, the wave function of the vibration mode of the mechanical resonator is $ \psi _{b} (t)=e^{i\chi _{b} (b^{\dag } b)^{2} t} {\left| \beta \right\rangle} $. For $ t=\pi /2\chi _{b} $, the wave function describes the superposition of two coherent states with opposite phases: $ \psi _{b} (\pi /2\chi _{b} )=\left[e^{i\pi /4} {\left| \beta \right\rangle} +e^{-i\pi /4} {\left| -\beta \right\rangle} \right]/\sqrt{2} $. Hence, the initial coherent state of the mechanical oscillator evolves, after a suitable amount of time has elapsed, into Yurke-Stoler-like states \cite{p42}. It is obviously that, due to symmetry of the Hamiltonian, the similar result is obtained in the case, when the mechanical oscillator is initially in the vacuum state and the cavity mode is in the coherent state $ {\left| \alpha \right\rangle} $. Consequently, the cavity wave function can be presented as $ \psi _{a} (\pi /2\chi _{a} )=\left[e^{i\pi /4} {\left| \alpha \right\rangle} +e^{-i\pi /4} {\left| -\alpha \right\rangle} \right]/\sqrt{2}$.

Now we also consider the case when the initial coherent state is realized for both the optical and mechanical oscillators, i.e. the density matrix of the system can be written as $ \rho ={\left| \alpha \right\rangle} {\left| \beta \right\rangle} {\left\langle \alpha \right|} {\left\langle \beta \right|} $. For example, for the optical oscillator, we obtain the following expression:
\[\rho _{a} (t)=e^{-\left|\alpha \right|^{2} } \sum _{n, m}^{\infty }\frac{\alpha ^{n} \alpha ^{*m} }{\sqrt{n!m!} } e^{i\chi _{a} (n_{a}^{2} -m_{a}^{2} )t}\times\]
\begin{equation}\label{eq12}
 \times\exp \left[\left|\alpha \right|^{2} (e^{-i\chi _{ab} (n_{a} -m_{a} )t} -1)\right]{\left| n_{a} \right\rangle} {\left\langle m_{a} \right|} .
\end{equation}

One can see that, due to the cross-Kerr interaction, the cavity and resonator modes are entangled during the evolution. However, at time $ t=2\pi /\chi _{ab} $
\[\rho _{a} (2\pi /\chi _{ab} )=e^{-\left|\alpha \right|^{2} } \sum _{n, m}^{\infty }\frac{\alpha ^{n} \alpha ^{*m} }{\sqrt{n!m!} }\times
\]
\begin{equation}\label{eq13}
\times \exp \left[i2\pi \frac{\chi _{a} }{\chi _{ab} } (n_{a}^{2} -m_{a}^{2} )\right]{\left| n_{a} \right\rangle} {\left\langle m_{a} \right|} .
\end{equation}

It follows from Eq. ~\ref{eq13} that at $ \chi _{a} /\chi _{ab} =1 $ the initial coherent state of the optical mode $ {\left| \alpha \right\rangle} $ is completely recovered at the mentioned moment. At $ \chi _{a} /\chi _{ab} =1/2 $, the state is transformed in $ {\left| -\alpha \right\rangle} $. At $ \chi _{a} /\chi _{ab} =1/4 $, the superposition state $ \left[e^{i\pi /4} {\left| \alpha \right\rangle} +e^{-i\pi /4} {\left| -\alpha \right\rangle} \right]/\sqrt{2} $ is realized and the density matrix of the system can be written as:
\[\rho (2\pi /\chi _{ab} )=\frac{1}{2} \left[{\left| \alpha \right\rangle} {\left\langle \alpha \right|} +{\left| -\alpha \right\rangle} {\left\langle -\alpha \right|}+\right.\]
\begin{equation}\label{eq14}
 \left.+i({\left| \alpha \right\rangle} {\left\langle -\alpha \right|} -{\left| -\alpha \right\rangle} {\left\langle \alpha \right|} \right]{\left| \beta \right\rangle} {\left\langle \beta \right|} .
\end{equation}

To estimate the coherence loss of the superposition coherent states due to their interaction with environment, Eqs. (8) for dissipators of the microwave cavity and the mechanical oscillator can be used. At small $t$ ($\kappa t \ll 1$ and $\gamma t\ll 1$), the coherence decay \cite{p43} is describes by the times $\tau_{a} \approx \left[2\kappa \left|\alpha \right|^{2} \right]^{-1}$ and $\tau _{b} \approx \left[2\gamma (2\bar{N}+1)\left|\beta \right|^{2} \right]^{-1} $ for the superposition states of the cavity and the mechanical oscillator, respectively. One can see that the decay times are strongly depend on the displacement parameters $\alpha $ and $\beta $ as well as the environment temperature via the mean thermal occupation $\bar{N}$. Using the following parameters: $\kappa /2\pi =242$ kHz, $\gamma /2\pi =228$ Hz, $\Omega /2\pi =36.2$ MHz, $\omega _{c} /2\pi =5.9$ GHz, $T=14$ mK \cite{p41}, and $\left|\alpha \right|^{2} =\left|\beta \right|^{2} =2$, we obtain that $\bar{N}\approx 7.6$, $\tau _{a} =1.64\times 10^{-7} $ s, $\tau _{b} =1.08\times 10^{-5} $ s, i.e. $\tau _{a}^{-1} /\kappa =4$ and $\tau _{b}^{-1} /\gamma =64.5$. So, the coherence decay rate of the photon superposition states in the microwave cavity is four times larger than their dissipation rate and for vibration quanta the coherence decay rate exceeds the dissipation rate by two orders of magnitude.

Thus, we have studied the optomechanical system with the asymmetric anharmonic mechanical resonator. When the frequencies of the mechanical oscillations exceed the parameters, characterizing the interaction between the optical and mechanical subsystems and their decay rates, as well as the values of cubic and quartic anharmonicities, this system can be replaced by an effective one using the Bogoliubov averaging method. We have found that in the Hamiltonian of the effective system, there are the Kerr interaction of photons in the optical cavity as well as the cross-Kerr interaction of photons and vibration quanta, induced by the oscillations in the asymmetric anharmonic potential of the mechanical resonator. In addition, the Kerr-like self-interaction of the mechanical resonator occurs. The bistable behavior of the number of vibration quanta as a function of the power of the mechanical driving and its detuning from the resonator frequency is predicted. This behavior is controlled by the cross-Kerr interaction. We have also shown that, in the absence of driving and dissipation, the constructed superposition Yurke-Stoler-like states of the cavity (or the mechanical resonator) disentangle at certain times the entangled modes of the system. Our approach can also be useful to study the hybrid system consisting of a quantum dot and a nanocavity mediated by a mechanical resonator \cite{p44, p45}. Further studies, both theoretical and experimental, would provide more insight to the behavior of optomechanical systems with asymmetric mechanical oscillations.


\begin{thebibliography}{44}
\bibitem{p1} F. Marquardt, J. P. Chen, A. A. Clerk, and S. M. Girvin, Phys. Rev. Lett. \textbf{99}, 93902 (2007).
\bibitem{p2} A. Schliesser, P. Del'Haye, N. Nooshi, K. J. Vahala, and T. J. Kippenberg, Phys. Rev. Lett. \textbf{97}, 243905 (2006).
\bibitem{p3} J. D. Teufel, J. W. Harlow, C. A. Regal, and K. W. Lehnert, Phys. Rev. Lett. \textbf{101}, 197203 (2008).
\bibitem{p4} F. Massel, T. T. Heikkil\"{a}, J.-M. Pirkkalainen, S. U. Cho, H. Saloniemi, P. J. Hakonen, and M. A. Sillanp\"{a}\"{a}, Nature \textbf{480}, 351 (2011).
\bibitem{p5} A.A. Clerk, F. Marquardt, and K. Jacobs, New J. Phys. \textbf{10}, 95010 (2008).
\bibitem{p6} M. Aspelmeyer, T.J. Kippenberg, and F. Marquardt, Rev. Mod. Phys. \textbf{86}, 1391 (2014).
\bibitem{p7} S. Mancini, V.I. Man'ko, and P. Tombesi, Phys. Rev. A \textbf{55}, 3042 (1997).
\bibitem{p8} X.-W. Xu, H. Wang, J. Zhang, and Y.-x. Liu, Phys. Rev. A \textbf{88}, 63819 (2013).
\bibitem{p9} M. Paternostro, D. Vitali, S. Gigan, M. S. Kim, C. Brukner, J. Eisert, and M. Aspelmeyer, Phys. Rev. Lett. \textbf{99}, 250401 (2007).
\bibitem{p10} D. Vitali, S. Gigan, A. Ferreira, H. R. B\"{o}hm, P. Tombesi, A. Guerreiro, V. Vedral, A. Zeilinger, and M. Aspelmeyer, Phys. Rev. Lett. \textbf{98}, 30405 (2007).
\bibitem{p11} L. Tian, Phys. Rev. Lett. \textbf{110}, 233602 (2013).
\bibitem{p12} L. Tian, Phys. Rev. Lett. \textbf{108}, 153604 (2012).
\bibitem{p13} Y.-D. Wang and A.A. Clerk, Phys. Rev. Lett. \textbf{108}, 153603 (2012).
\bibitem{p14} J. Bochmann, A. Vainsencher, D. D. Awschalom, and A. N. Cleland, Nature Phys \textbf{9}, 712 (2013).
\bibitem{p15} G.S. Agarwal and S. Huang, Phys. Rev. A \textbf{81}, 041803 (2010).
\bibitem{p16} H. Jing, \c{S}.K. \"{O}zdemir, Z. Geng, J. Zhang, X.-Y. L\"{u}, B. Peng, L. Yang, and F. Nori, Scientific reports \textbf{5}, 9663 (2015).
\bibitem{p17} P.-C. Ma, J.-Q. Zhang, Y. Xiao, M. Feng, and Z.-M. Zhang, Phys. Rev. A \textbf{90}, 43825 (2014).
\bibitem{p18} J.M. Dobrindt, I. Wilson-Rae, and T.J. Kippenberg, Phys. Rev. Lett. \textbf{101}, 263602 (2008).
\bibitem{p19} S. Huang and G.S. Agarwal, Phys. Rev. A \textbf{80}, 033807 (2009).
\bibitem{p20} T. T. Heikkil\"{a}, F. Massel, J. Tuorila, R. Khan, and M. A. Sillanp\"{a}\"{a}, Phys. Rev. Lett. \textbf{112}, 203603 (2014).
\bibitem{p21} M. Bhattacharya and P. Meystre, Phys. Rev. Lett. \textbf{99}, 073601 (2007).
\bibitem{p22} A. Xuereb and M. Paternostro, Phys. Rev. A \textbf{87}, 023830 (2013).
\bibitem{p23} J.D. Thompson, B.M. Zwickl, A.M. Jayich et al., Nature \textbf{452}, 72 (2008).
\bibitem{p24} Y.R. Shen, \textit{The principles of nonlinear optics} (Wiley-Interscience, New York, Hoboken, 2003).
\bibitem{p25} Z. R. Gong, H. Ian, Y.-x. Liu, C. P. Sun, and F. Nori, Phys. Rev. A \textbf{80}, 065801 (2009).
\bibitem{p26} P. Rabl, Phys. Rev. Lett. \textbf{107}, 063601 (2011).
\bibitem{p27} A. Nunnenkamp, K. B{\o}rkje, and S.M. Girvin, Phys. Rev. Lett. \textbf{107}, 063602 (2011).
\bibitem{p28} V. Kaajakari, T. Mattila, A. Oja et al., J. Microelectromech. Syst. \textbf{13}, 715 (2004).
\bibitem{p29} P. Huang, J. Zhou, L. Zhang, D. Hou, S. Lin, W. Deng, C. Meng, C. Duan, C. Ju, X. Zheng, F. Xue, and J. Du, Nature communications \textbf{7}, 11517 (2016).
\bibitem{p30} K. Jacobs and A.J. Landahl, Phys. Rev. Lett. \textbf{103}, 067201 (2009).
\bibitem{p31} J.-Q. Zhang, Y. Li, M. Feng et al., Phys. Rev. A \textbf{86}, 053806 (2012).
\bibitem{p32} C. Kong, H. Xiong, and Y. Wu, Phys. Rev. A \textbf{95}, 033820 (2017).
\bibitem{p33} Z. Wu, R.-H. Luo, J.-Q. Zhang, Y.-H. Wang, W. Yang, and M. Feng, Phys. Rev. A \textbf{96}, 033832 (2017).
\bibitem{p34} S. Huang, H. Hao, and A. Chen, Applied Sciences \textbf{10}, 5719 (2020).
\bibitem{p35} X.-Y. L\"{u}, J.-Q. Liao, L. Tian, and F. Nori, Phys. Rev. A \textbf{91}, 013834 (2015).
\bibitem{p36} L. Latmiral, F. Armata, M. G. Genoni, I. Pikovski, and M. S. Kim, Phys. Rev. A \textbf{93}, 052306 (2016).
\bibitem{p37} N.N. Bogoliubov and Y.A. Mitropolsky, \textit{ Asymptotic Methods in the Theory of Nonlinear Oscillations} (Gordon and Breach, New York, 1961).
\bibitem{p38} A.P. Saiko, S.A. Markevich, and R. Fedaruk, Phys. Rev. A \textbf{93}, 063834 (2016).
\bibitem{p39} A.P. Saiko, S.A. Markevich, and R. Fedaruk, Phys. Rev. A \textbf{98}, 043814 (2018).
\bibitem{p40} H.J. Carmichael, \textit{ Statistical Methods in Quantum Optics 2: Non-Classical Fields} (Springer-Verlag, Berlin, Heidelberg, 2008).
\bibitem{p41} V. Singh, O. Shevchuk, Y.M. Blanter, and G.A. Steele, Phys. Rev. B \textbf{93}, 245407 (2016).
\bibitem{p42} B. Yurke and D. Stoler, Phys. Rev. Lett. \textbf{57}, 13 (1986).
\bibitem{p43} H. Saito and H. Hyuga, J. Phys. Soc. Jpn. \textbf{65}, 1648 (1996).
\bibitem{p44} Z.-L. Xiang, S. Ashhab, J. Q. You, and F. Nori, Rev. Mod. Phys. \textbf{85}, 623 (2013).
\bibitem{p45} J.E. Ram\`{\i}rez-Mun\={o}z, J.P. Restrepo Cuartas, and H. Vinck-Posada, Phys. Lett. A \textbf{382}, 3109 (2018).
\end{thebibliography}
\end{document}